**Quantum Algorithm Processors to Reveal Hamiltonian Cycles**
Professor J. R. Burger

*Summary* – Quantum computer versus quantum algorithm processor in CMOS are compared to find (in parallel) all Hamiltonian cycles in a graph with m edges and n vertices, each represented by k bits. A quantum computer uses quantum states analogous to CMOS registers. With efficient initialization, number of CMOS registers is proportional to (n-1)! Number of qubits in a quantum computer is approximately proportional to kn+2mn in the approach below. Using CMOS, the bits per register is about proportional to kn, which is less since bits can be irreversibly reset. In either concept, number of gates, or operations to identify Hamiltonian cycles is proportional to kmn. However, a quantum computer needs an additional exponentially large number of operations to accomplish a probabilistic readout. In contrast, CMOS is deterministic and readout is comparable to ordinary memory.

**Introduction**
The problem of searching for Hamiltonian cycles, or circuits in a given graph is known to be NP-complete, that is, nondeterminant in a classical computer, and always polynomial in a massively parallel processor. Ability to solve such problems exactly might benefit many areas, including the layout of integrated circuits [1,2]. In a weighted graph, exact solution to the famous salesman's problem may be found as a Hamiltonian circuit with minimum total weight [3]. Once all Hamiltonian circuits are identified, it is easy to calculate weight using an ordinary computer to choose the minimum.

A typical algorithm employs a strategy known as depth-first search [4]. It works well for smaller numbers of vertices, but grows exponentially with n, the number of vertices. It assumes a directed graph. This paper uses the more general undirected graph in which paths may go either way on an edge.

A wiring diagram is given below for reversible processing as required in a quantum computer. Alternately, combinations of vertices may be analyzed using the author's novel quantum algorithm processor in CMOS [5]. CMOS is deterministic. Data outputs may be observed with full probability at any point in the system. Also, CMOS can be nonreversible to simplify wiring diagrams. This feature was demonstrated when finding all divisors of an integer [6]. That paper, like this one is mainly to show the power of a quantum algorithm processor in CMOS, available to any engineer clever enough to build one for solving real (non academic) problems.

**Problem Definition** -- Let G = (V, A) be a graph in which V = $\{v_1, v_2, ...v_n\}$ is the set of n vertices, and A is the set of m arcs $(v_i, v_j)$. A Hamiltonian cycle (or circuit) in G is a permutation $(s_i)$ of the vertices such that $(v_s, v_{s+1})$ belongs to A for i = 1, 2, ..., n-1. Also $(v_{sn}, v_{s1})$ must belong to A to close the circuit.

**Solution Approach** – To communicate the method, consider Figure B1a.

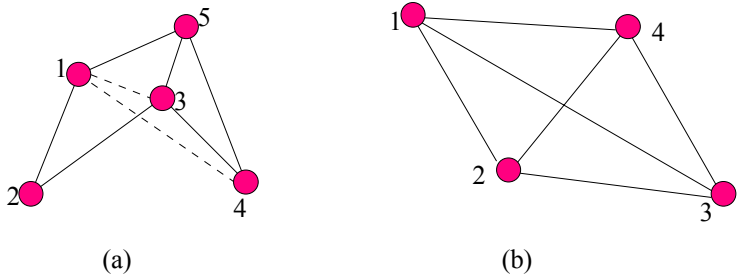

(a)                              (b)

**Figure B1   Examples of Hamiltonian circuits**

A Hamiltonian circuit is 1-2-3-4-5-1.  Examples of circuits that are not Hamiltonian are (1-2-3-5-1) and (1-2-3-5-4-5-1) because the first leaves out node 4 while the second visits node 5 twice.  Figure B1b is characterized by several Hamiltonian circuits: 1-2-3-4-1; 1-2-4-3-1; 1-3-2-4-1; each can be walked backwards, for example- 1-4-3-2-1.

Nielson and Chuang began the process of finding Hamiltonian circuits using a quantum computer [2].  One approach to finding Hamiltonian circuits is to consider all permutations of vertices.  Each permutation refers to a path within the graph.  The existence of an edge must be checked for each pair of vertices.  Once a permutation of vertices is proposed, it must be verified somehow that no vertex is visited more than once, and that all are visited.  Finally, the path must end at the vertex where it began, vertex 1 when CMOS is used.

In a reversible system of logic appropriate to a quantum computer, or a quantum algorithm processor in CMOS, a wiring diagram is the appropriate tool.  The wiring diagram tells what happens to the bits in each register in parallel.  Built into the diagram is that the correct number of vertices are touched.  Figure B2 illustrates Hamiltonian circuit recognizers.

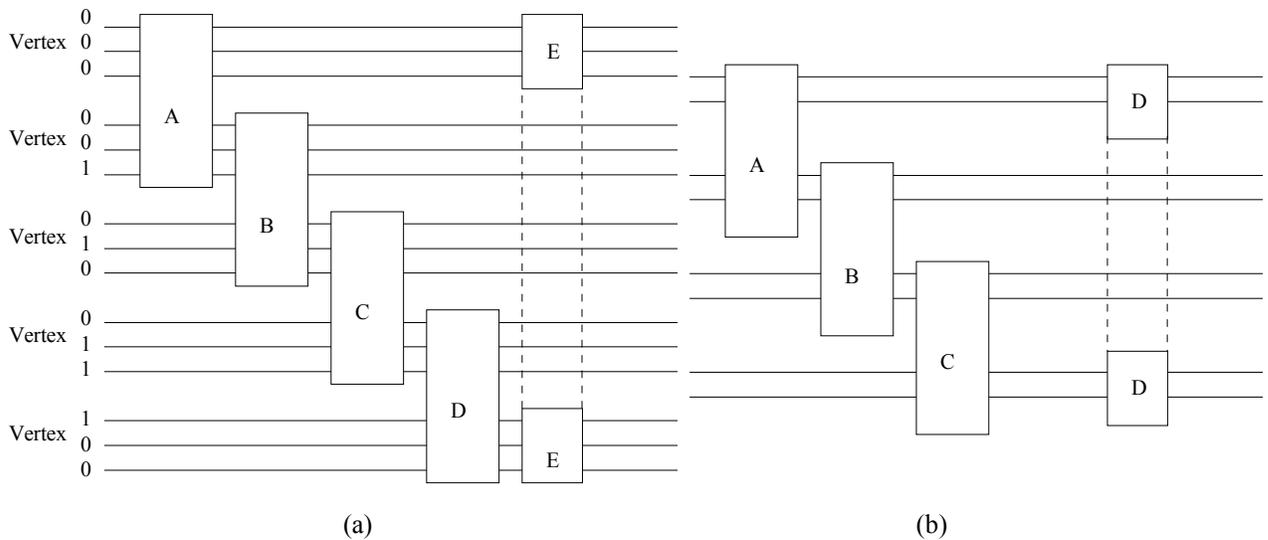

(a)                  (b)

**Figure B2  Hamiltonian circuit recognizers (Workspace lines not shown)**

Vertices are numbered from 0 to n-1; k bits represent the vertex numbers. If n is a power of two, $k = Log_2(n)$. There are kn wires. The vertically organized binary number on the left of the figure denotes the initial condition of the register. The initialization indicated above is the sequence 0,1,2,3,4, that is, nodes 1-2-3-4-5 in Figure B1a. CMOS initialization of course varies for each register in a quantum algorithm processor.

The boxes in the figure detect the existence of an edge between the two vertices as specified by the input code. If all boxes (A-E) detect edges, a Hamiltonian cycle might exist. In a CMOS implementation only candidate circuits are tested. For example, beginning form vertex 1, the first set of choices for a vertex are limited to n-1, since an edge cannot return to where it began. The second choice can be limited to n-2, since not only can it not return to where it began, it cannot re-visit a previously visited vertex. Assuming no vertex is visited twice, the number of initializations is (n-1)(n-2)….(1) or (n-1)! Note that the last vertex must be one that goes to vertex 1 to complete the circuit. It is assumed below that the number (NUM) of candidate circuits in a CMOS implementation is:
NUM = (n-1)!

The generation of initializations can be done in an ordinary computer with reference to a matrix of edges (Refer to the 5-vertex example in Table 1 below).

Table 1

|    | 12 | 13 | 14 | 15 |
|----|----|----|----|----|
| 21 |    | 23 | 24 | 25 |
| 31 | 32 |    | 34 | 35 |
| 41 | 42 | 43 |    | 45 |
| 51 | 52 | 53 | 54 |    |

A circuit begins at the top. Only one entry is allowed in a given row and column. After the fourth edge a jump back to vertex 1 is assumed. Examples are:

**12**-23-34-45-51
**12**-23-35-54-41
**12**-24-45-53-31

To get all Hamiltonian circuits, it is necessary to test each of the (n-1)! initializations for the existence of a circuit.

It is simpler of course to process all permutations, $2^{nk}$, that is, $2^{n \, Log(n)} = n^n$, which is exactly what a quantum computer would do. That is, it processes all states corresponding to all binary counts using nLog(n) bits. But this takes an excessive number of CMOS registers. For example, if n = 10, then $2^{n \, Log(n)} = 10,000,000,000$, in other words, nLog(n) ≈ 34 qubits (plus extra qubits for workspace). On the other hand, CMOS requires (n-1)! = 362,880 registers, which is insignificant in this age of gigabyte memories.

If all permutations are tried, there are other issues. Bad cycles can appear as Hamiltonian circuits. For example, 1-2-3-1-4 might crop up in Figure B1a, or 1-2-1-2 might crop up in Figure B1b. Bad cycles that do not reuse an edge are even more likely in a larger graph. To overcome this problem, edges must be disabled if they come from or go to a vertex that has been visited. In effect, once a vertex is used, it will not be used again, somewhat like above when choosing initializations. The lone exception will be the $n^{th}$ vertex, for example, the one processed by box E in Figure B2a. The $n^{th}$ vertex will not be subject to enabling, but must return to the beginning vertex, vertex one, any way it can. A quantum computer needs the option of using all permutations, should technology permit. Later, during optimization for CMOS, extra lines are easy to remove.

The programming of an edge detector follows from the given description of the graph. Figure B3 provides a basic example using Figure B1a.

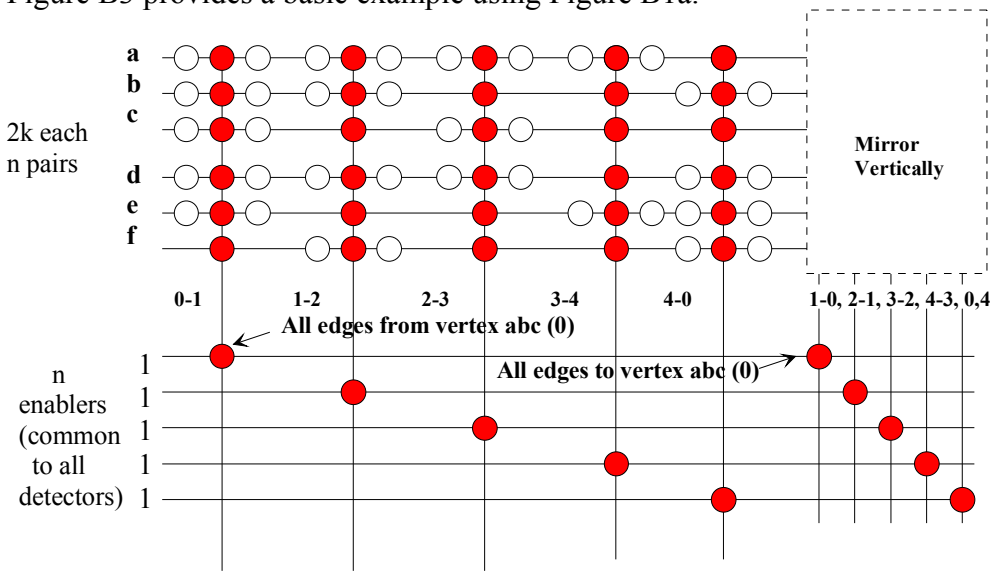

**Figure B3   Edge Detection**

The abc,def entering from the left represents binary code for a given source-destination vertex. Each column of dots represents a physical edge, one of m, as specified in the edge description. For example, if the edge 0,1 is present, the code 000,001 is first converted to all ones as shown. The corresponding enable line for edges from vertex 0 also has to be true. Subsequently a Controlled Not (7CN) is executed. The 7CN flips a result bit only if all six inputs and the enable are true.

Figure B4 illustrates how a 7CN is implemented using Double Controlled Nots (DCN). If inputs are true, the result line goes true, which implies the existence of an edge. Without difficulty the 7CN can be redesigned to a 2k+1-Controlled Not, using 2k lines.

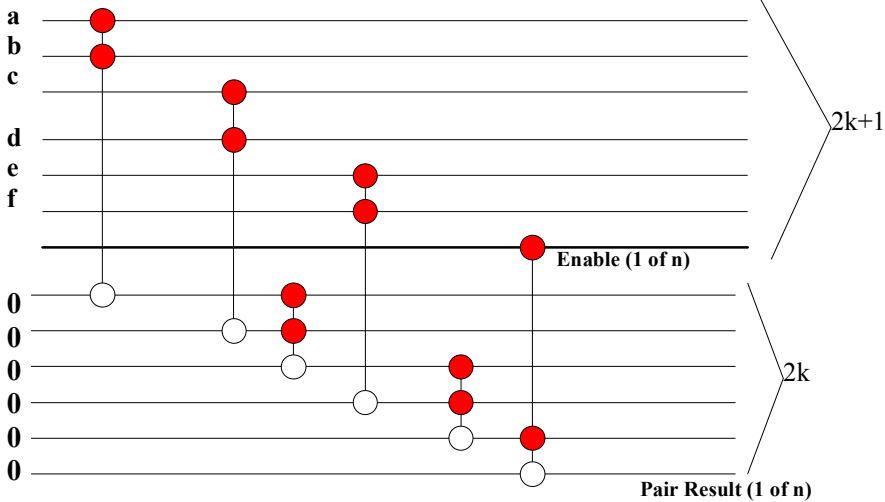

**Figure B4   Pair result**

It was described how edges such as 0,1 are detected. In a undirected graph, it is also necessary to detect edges such as 1,0 going the other way. The easiest way is to mirror the above gates vertically, in order to detect edges going the other way. The box in Figure B3 shows this. As shown, the detector registers either ascending or descending edge coordinates. There is an edge detector for each of the 2m edges, for each of the n pairs.

**Reversible Latching**
If all possible permutations are presented, as they would in a quantum computer, the enablers above are necessary to prevent the same vertex from being visited more than once. Disabling a vertex is accomplished by disabling all edges to or from a given vertex as in Figure B5.

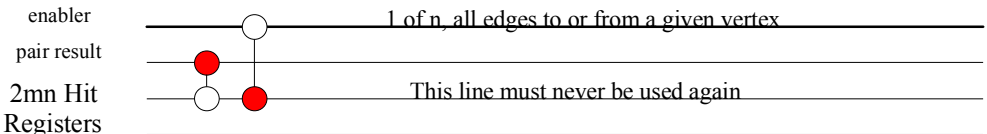

**Figure B5   Vertex disable method**

There is a pair result line for each of the n pairs. If a pair result line goes true, it means that an edge exists for that particular pair of vertices. Once it goes true, it stays true, because no other edge will fit that pair of vertices. The pair result line cannot be used directly to toggle an enabler line, since it would re-toggle at each subsequent edge detector. To record the fact that a edge from a given vertex (or to a given vertex) has been used, a dedicated line (or hit register) serves to remember and to disable the appropriate enable line. There are m edges with ascending vertices, and m with descending vertices, and n pairs; the number of hit registers is 2mn.

The above system latches the enabler line to zero. It is not claimed to be the only way to do this, but at least it is certain to keep the enabler latched to zero, since unwanted toggling cannot be tolerated.

**Restoring Scratch Pad Lines**
The 2k-1 lines used for scratch pad above can be used again if they are reset to zero after each usage, as in Figure B6.

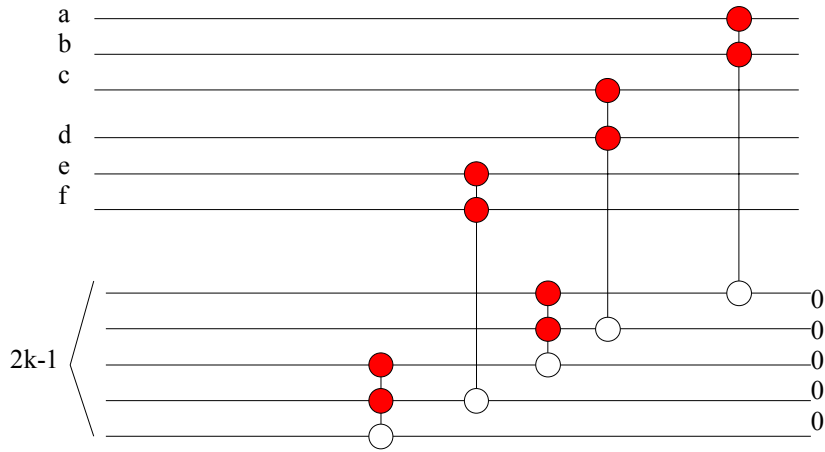

**Figure B6   Restoring scratch pad lines**

It is important to note that the $n^{th}$ pair will not involve the enable lines, because the circuit must go again to the first vertex in a Hamiltonian circuit.

**Closing the Circuit**
Each pair has a result line as in Figure B7.

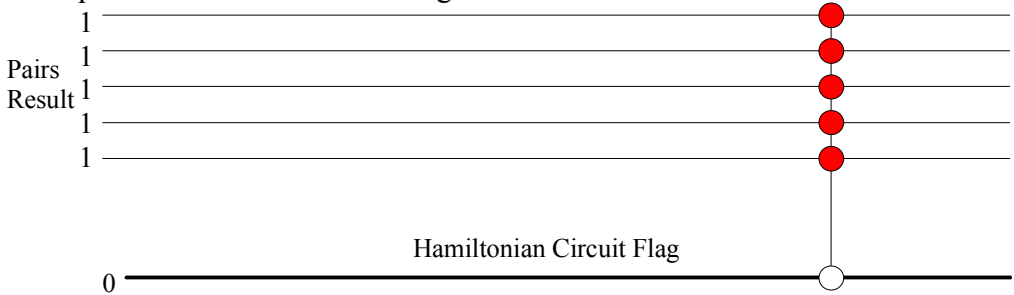

**Figure B7   Final Result**

There are n pairs. If all pair results are true for a particular input code, it means that enough edges have been found to form a closed circuit. A Hamiltonian circuit is flagged by detecting all ones using a method like that in Figure B4. This requires n-1 scratch pad lines, or n-1-(2k-1) = n–2k additional lines for scratch pad purposes.

**Performance Analysis**
The number of registers (NUM) using CMOS is discussed above. The number of lines or bits per register (BITS) can be counted approximately as follows: Basic workspace = nk, Pairs result lines = n, Enables = n, Controlled NOT scratch = n-1, Hit lines = 2nm.
BITS = nk + n + n + n-1 + 2nm
= n(k+2m+3) - 1
$\approx$ n(k+2m)
This is reduced in a CMOS implementation by eliminating the enable and hit lines below.

**Time to complete** – Operations (OPS) can be accounted approximately as: Edge detection = 3(2m)n, Controlled NOT pair result = 4k(2m)n, Hits = 2(2m)n, Final result = n-1, summarized as
OPS = 6mn + 8kmn + 4mn + n - 1
= n[m(8k+10)+1]-1
$\approx$ 8kmn

This value approximates the number of gates, or time slots required. The main delay is from the Controlled NOT implementation.

**Taking advantage of irreversible logic**
A feature of CMOS technology is that, aside from being deterministic, it is capable of being irreversible. For the purpose of this paper, being irreversible means that a bit can be reset to zero. Symbols for this are suggested in Figure B8.

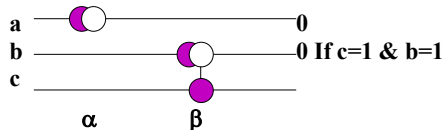

**Figure B8 Irreversible operations available**

At $\alpha$ the a-line is zeroed if a = 1. If a = 0, it stays zero, so it is unconditionally reset to zero. At $\beta$ the b-line is zeroed only if c=1 while b=1. If b=0, it stays zero. If c=0, b is unchanged. These are irreversible, since for example, once zeroed, the original value of line 'a' cannot be reconstructed. Operations as above are readily available in the author's quantum algorithm processor using CMOS. Algorithms that use zeroing go well beyond quantum computers, since quantum systems are supposed to be reversible.

Irreversible logic may be used to advantage in the one-shot enable method in Figure B9.

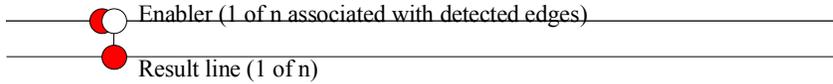

**Figure B9   Resetting enable to zero**

This saves 2mn lines that otherwise must serve as hit registers. The reduced number of bits, or lines is:
BITS = nk + n + n + n-1
= n(k+3) - 1

However, once it is realized that enables are unnecessary, owing to the structuring of the initializations, this reduces to:
BITS = n(k+2) - 1
$\approx$ kn

Time to complete is dominated by scratch pad operations and is about the same, except time for hits is halved (4mn be omes 2mn):
OPS = 6mn + 8kmn + 2mn + n - 1
$\approx$ 8kmn

Since CMOS clocks can be fast [6], number of operations is not expected to be as critical as the number of registers n!/n and bits per register kn. These two parameters define the size of the required integration expressed as total bits (TOTBITS). The TOTBITS product above is about:
TOTBITS $\approx$ k n!

Figure B10 indicates the total number of bits that a quantum algorithm processor in CMOS must reserve to find all Hamiltonian circuits using n vertices. This quantity is large because of the (n-1)! registers. This number can be reduced by using information about the location of edges when calculating initializations (Not done here).

**Observing Outputs**
The Hamiltonian circuit flag bit exists physically in each of the (n-1)! CMOS registers. A simple integrated OR-function of the flag bits would indicate immediately whether or not a result exists. Actually finding the Hamiltonian circuit definitions, that is, the permutation of n vertices associated with true flags is more of a challenge. Although details about readout circuitry are beyond the scope of this paper, in general there will have to be output pins, each with 'trees' that extend to (n-1)! flags. The outputs would indicate the locations of the true flags. Particular permutations of n vertices, that is, initializations, would be available in an external database, stored in order.

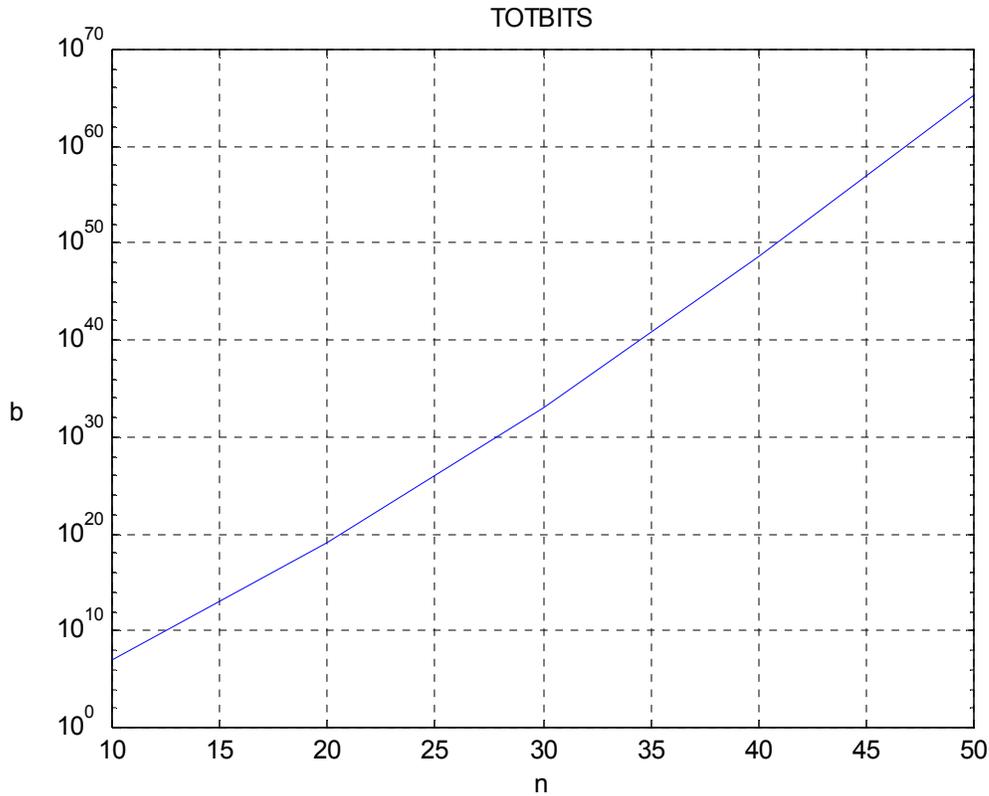

Figure B10   Bits required (b)

It is unnecessary to read particular permutations, although this is an option if conventional addressing is employed for the memory registers. In content addressable memory systems it is commonplace to perform a multi read, in which data for true flags are read one at a time. Unfortunately, if a very large number of flags are true, electrical output is time consuming and impractical.

It might be possible to locate true flags using a laser beam, but this departs from CMOS technology.

In contrast to integrated circuit thinking, output for a quantum computer is even more visionary, and then with only a certain chance of getting a correct answer, as concluded by recent work [2]. Defined above is a binary function for which certain inputs makes a result line go true. By initializing the result line to ½ ( |0> - |1>) as is understood in quantum computer parlance, Hamiltonian circuits within $n^n$ permutations become tagged with a negative sign. The permutations can be identified using Grover iteration, which requires a massive $\sqrt{n^n}$ steps approximately. Grover iteration greatly increase the probability that a correct answer will be read.   $\sqrt{n^n}$ steps approximately must done for each Hamiltonian circuit, so if there are a large number of Hamiltonian circuits approaching (n-1)! output is bound to be time-consuming and impractical, same as CMOS.

**Conclusions**

Presented above is a comparison of CMOS versus quantum computer concepts for revealing Hamiltonian cycles. By taking advantage of its deterministic and nonreversible nature, CMOS simplifies. After reductions, there are about kn bits per register. The number of gates, proportional to completion time, is approximately 8kmn. The most general search used (n-1)! registers operating in parallel.

Following the line of thought that led to (n-1)! registers in which a vertex is discarded once visited, a reversible quantum computer would require about kn + 2mn qubits in the author's programming. The number of gates, proportional to completion time, remains at approximately 8kmn in order to set flags. Still remaining is the quantum search algorithm of order $2^{n\text{Log}(n)/2}$ operations for each Hamiltonian circuit. A remaining puzzle is that a quantum computer using the above programming will recognize the same Hamiltonian circuit beginning from each of n vertices, and either clockwise or counter clockwise. This gives 2n redundant circuits.

Obviously, if a classical computer had to be used, there would be an exponential slowdown. A method is provided above to program up to (n-1)! little classical computers, or registers in parallel using technology that is available to anyone who wants to solve hard problems. All Hamiltonian circuits can be found in an arbitrary graph if desired. Given the fact that integrated circuit technology is growing exponentially, that is, larger integrations, and smaller, faster circuits by factors of two each year, there is no doubt in the author's mind that quantum algorithm processing in CMOS is a real force.